\documentclass[fleqn,usenatbib]{mnras}

\usepackage{mathptmx}

\usepackage[T1]{fontenc}
\usepackage[utf8]{inputenc}
\usepackage{ae,aecompl}

\usepackage{graphicx}	
\usepackage{amsmath}	
\usepackage{amssymb}	

\title[Fornax GC timing problem ]{Cusp or core? Revisiting the
  globular cluster  timing problem in Fornax}

\author[Meadows et al.]{
\parbox[t]{\textwidth}{
Noah Meadows$^{1}$, Julio F. Navarro$^{1}$, Isabel Santos-Santos$^1$, Alejandro
Ben\'itez-Llambay$^2$, Carlos Frenk$^2$
}
\\
\\
$^{1}$Department of Physics and Astronomy, University of Victoria, Victoria, BC V8P 5C2, Canada\\
$^{2}$Institute for Computational Cosmology, Durham University, South Road, Durham DH1 3LE, United Kingdom\\
}

\pubyear{2019}

\begin{document}
\label{firstpage}
\pagerange{\pageref{firstpage}--\pageref{lastpage}}
\maketitle

\begin{abstract}
  We use N-body simulations to revisit the globular cluster (GC)
  ``timing problem'' in the Fornax dwarf spheroidal (dSph).  In
  agreement with earlier work, we find that, due to dynamical
  friction, GCs sink to the center of dark matter halos with a cuspy
  inner density profile but ``stall'' at roughly 1/3 of the core
  radius ($r_{\rm core}$) in halos with constant-density cores. The
  timescales to sink or stall depend strongly on the mass of the GC
  and on the initial orbital radius, but are essentially the same for
  either cuspy (NFW) or cored halos normalized to have the same total
  mass within $r_{\rm core}$. Arguing against a cusp on the basis that
  GCs have not sunk to the center is thus no different from arguing
  against a core, unless all clusters are today at
  $\sim (1/3)\, r_{\rm core}$. This would imply a core radius
  exceeding $\sim 3$ kpc, much larger than seems plausible in any
  core-formation scenario. (The average projected distance of Fornax
  GCs is $\langle R_{\rm GC,Fnx}\rangle\sim 1$ kpc and its effective
  radius is $\sim 700$ pc.)  A simpler explanation is that Fornax GCs
  have only been modestly affected by dynamical friction, as expected
  if clusters started orbiting at initial radii of order $\sim 1$-$2$
  kpc, just outside Fornax's present-day half-light radius but well
  within the tidal radius imprinted by Galactic tides. This is not
  entirely unexpected. Fornax GCs are significantly older and more
  metal-poor than most Fornax stars, and such populations in dSphs
  tend to be more spatially extended than their younger and more
  metal-rich counterparts. Contrary to some earlier claims, our
  simulations further suggest that GCs do not truly ``stall'' at
  $\sim 0.3\, r_{\rm core}$, but rather continue decaying toward the
  center, albeit at reduced rates.  We conclude that dismissing the
  presence of a cusp in Fornax based on the spatial distribution of
  its GC population is unwarranted.
\end{abstract}




\section{Introduction}
\label{sec:intro}

The globular cluster (GC) system of the Fornax dwarf spheroidal (dSph)
is often cited as evidence for the presence of a constant-density core in the dark
matter halo density profile. The issue has been addressed repeatedly
in the literature, starting with the early work of
\citet{Hernandez1998}, who were among the first to describe how the
spatial distribution of globular clusters may be used to gain insight
into the dark matter density distribution in dSphs. This elaborated on
the earlier work of \citet{Tremaine1976}, who puzzled about the lack
of a central stellar ``nucleus'' in Fornax, expected from the orbital
decay and subsequent fusion of its GCs. Indeed, the $5$ GCs in Fornax
are widely spread through the galaxy, with an average projected
radius\footnote{For comparison, Fornax's effective radius is
  $R_{\rm eff,Fnx} \sim 700$ pc \citep{Irwin1995}.} of
$\langle R_{\rm GC,Fnx} \rangle \sim 1$ kpc \citep{Mackey2003a},
despite the fact that their orbital decay timescales, inferred at the
time from simple analytical dynamical friction estimates
\citep{Chandrasekhar1943}, were substantially shorter than their ages.

This puzzle is widely referred to as the Fornax ``GC timing problem''
and has elicited the proposal of a number of possible solutions,
ranging from the ``dynamical stirring'' of GC orbits by Galactic tides or massive
black holes \citep{Oh2000}, to more straightforward options, such as
assuming that GCs in Fornax started decaying from initial radii
somewhat larger than where they are currently at
\citep{Angus2009, Boldrini2019}.

An alternative solution was proposed by
\citet{Goerdt2006}, who reported some of the first fully
self-consistent N-body simulations of the problem. These authors
found that analytical predictions for dynamical friction-induced
orbital decay fail in the case of halos with constant-density
cores. Instead of continually decaying, GCs ``stall'' once they are
well inside the core, at a radius that is roughly independent of 
GC mass. In cuspy halos, such as the Navarro-Frenk-White profiles
of cold dark matter (CDM) halos
\citep[NFW;][]{Navarro1996a,Navarro1997}, GCs do not stall but rather
sink until they reach either the center or a radius where the
enclosed dark mass is comparable to that of the cluster
\citep{Goerdt2010}.

The ``stalling radius'' result has been reproduced in subsequent
 work \citep[see; e.g.,][]{Read2006,Inoue2009,Petts2015,Kaur2018}, and
has become an often cited argument for the presence of a core in
Fornax: if GCs ``stall at the core radius'', as is often claimed, then for
$\langle R_{\rm GC,Fnx} \rangle \approx r_{\rm core}\sim 1$ kpc the
timing problem would be solved.

A core radius of that size would be comparable to Fornax's effective
radius, as expected if cores are carved out of cuspy, NFW halos by
baryonic inflows/outflows during the formation of the galaxy
\citep[see; e.g.,][and references
therein]{Navarro1996b,Pontzen2012,DiCintio2014}. It would also be
commesurate with  the core size expected for Fornax in models
where cores are produced by ``self-interactions'' between dark matter
particles \citep{Spergel2000,Rocha2013,Kaplinghat2016}, at least for
self-interacting cross sections in the preferred range of $0.1$-$1$
cm$^2$/g. These coincidences have helped galvanize support for the ``core''
solution to the Fornax GC timing problem.

One problem with this solution is that the stalling radius is actually
well inside the core\footnote{We shall hereafter define the core
  radius, $r_{\rm core}$, as the (3D) radius where the dark matter
  density drops by a factor of two from its central value. Since this
  convention is not always followed, care is needed when comparing
 quantitative  results from different authors.}; i.e.,
$r_{\rm stall}\sim 0.3 \, r_{\rm core}$. Taken at face value, this
would imply that a core radius as large as $\sim 3$ kpc would be
needed to solve the timing problem, a value that seems, in principle,
much larger than can be reasonably accommodated by current
core-formation models.

One reason why cores remain a viable solution is that subsequent
simulation work uncovered a rather puzzling phenomenon that affects
clusters that reach the inner regions of the core. In the simulations
reported by \citet{Cole2012}, clusters well inside the core tend to
{\it gain} orbital energy, and are pushed out by ``dynamical
buoyancy'', a mechanism whose detailed origin remains unclear but
which apparently counteracts dynamical friction in the
innermost regions. The combination of friction and buoyancy could, in
principle, lead to a stationary ``shell-like'' distribution of
globulars near the core radius, where the two effects would presumably
cancel out. Although appealing, this result relies on a mechanism that
is still poorly understood and that urgently needs theoretical
underpinning and independent numerical confirmation.

We address some of these issues here using a series of N-body simulations of
the decay of GCs in cuspy or cored halos. We
focus on the difference in the timescales needed for clusters to
``sink'' (i.e., to reach the center, in the case of cusps), or to ``stall''
(in the case of cores). We also follow the long-term evolution of
several clusters after they stall, in order to learn about the possible
effects of dynamical buoyancy on these systems.

This paper is organized as follows. Sec.~\ref{SecNumSims} describes
our numerical setup, while our main results are presented in
Sec.~\ref{SecRes}. We conclude with a discussion of the applicability
of these results to Fornax and to the ongoing cusp vs core debate in
Sec.~\ref{SecConc}.

\begin{figure}
\includegraphics[width=\columnwidth]{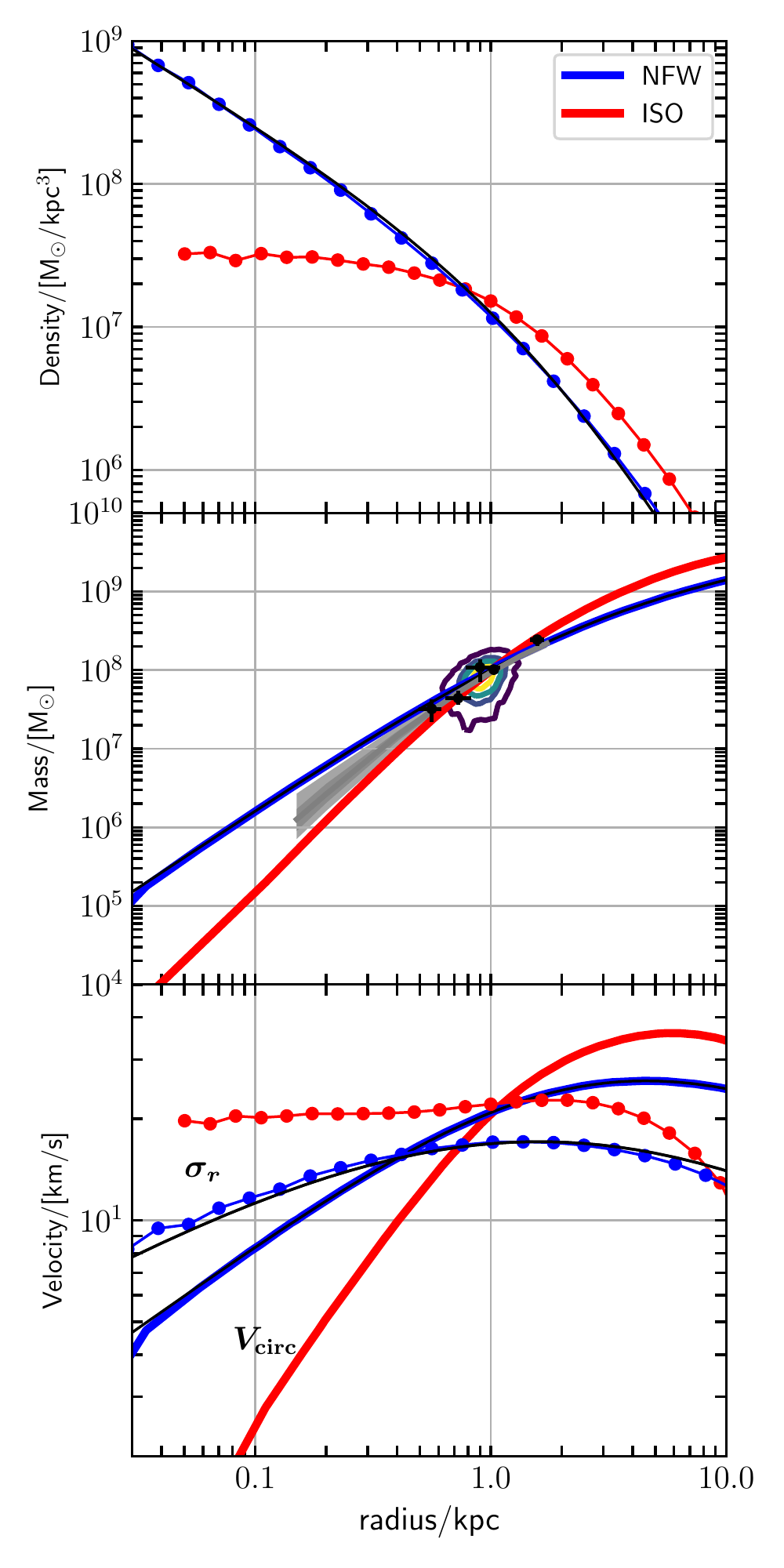}
\caption{Density (top), enclosed mass (middle) and circular
  velocity/radial velocity dispersion
  (bottom) profiles of the halo models used in this study. The
  profiles (shown with circles/thick curves) correspond to the
  16M-particle N-body realization of each model, and are plotted after
  the halo has been run for $\sim 4$ Gyr to allow it to relax to
  equilibrium. Blue corresponds to the cuspy NFW halo, and red to the
  non-singular isothermal (cored) halo. The analytic NFW profile is
  shown with thin black lines. The contours in the middle panel are
  constraints on the enclosed mass within $\sim 1$ kpc, derived from
  the stellar velocity dispersion and density profiles \citep[see][for
  details]{Fattahi2016}. In the same panel, crosses indicate the
  estimates of \citet{Walker2011} and \citet{Amorisco2013}. The grey
  shaded band corresponds to the recent kinematic analysis of
  \citet{Read2019}. All of these estimates coincide at $r\sim 1$ kpc.}
    \label{FigHalos}
\end{figure}

\section{Numerical Simulations}
\label{SecNumSims}

The simulations follow the evolution of a GC (represented by a softened point mass)
in  two spherical N-body halo models.  The first
model is a cuspy, NFW halo (hereafter, ``NFW'') with parameters
consistent with those expected in a Planck-normalized $\Lambda$CDM
cosmology \citep{Ludlow2016}. The second model is a non-singular
isothermal sphere (hereater, ``ISO'') normalized to have the same mass
as the NFW profile within its core radius.

\subsection{Halo models}
\label{SecHaloModels}

The cuspy halo model follows an NFW profile,
\begin{equation}
  \rho(r)={\rho_s \over (r/r_s)(1+r/r_s)^2},
\end{equation}
and is fully specified by two parameters; e.g., a scale
density, $\rho_s$, and a scale radius, $r_s$, or, alternatively, a
maximum circular velocity, $V_{\rm max}$ and the radius at which it is
achieved, $r_{\rm max}$. The two radial scales are related by
$r_{\rm max}=2.16\, r_s$.

The cored halo is modeled as a non-singular isothermal sphere
\citep[see; e.g.,][p.228]{BinneyTremaine1987}. Although there is no
simple algebraic formula to describe this model, it is also fully
specified by two parameters, usually expressed as the central density,
$\rho_0$, and the core radius, $r_{\rm core}$. To prevent divergences,
the models are truncated with an exponential taper in the outer
regions, but this should be of little consequence for our analysis.

The models are assumed to have isotropic velocity distributions and
are normalized to have the same enclosed mass within the deprojected
(3D) half-light radius of Fornax, $M(<1$ kpc$)=10^8\, M_\odot$,
inferred from observations of the line-of-sight velocity dispersion
and projected light profile of Fornax
\citep{Walker2009,Wolf2010}. This is widely agreed to be the most
robust dynamical mass estimate available for this system
\citep[see the discussion of Fig.~1 in][and references therein]{Fattahi2016}.

Fig.~\ref{FigHalos} contrasts the density, $\rho(r)$, circular
velocity, $V_c(r)$, enclosed mass, $M(r)$, and radial velocity
dispersion, $\sigma_r(r)$, profiles of the two models. The NFW profile
has $r_s=2.11$ kpc and
$\rho(r_s)=(\rho_s/4)=3\times 10^6\, M_\odot/$kpc$^3$. This
corresponds to a ``virial''\footnote{Virial quantities are
  conventionally defined as those measured at a radius where the mean
  enclosed density equals $200\times$ the critical density for
  closure, and are identified with a ``200'' subscript.} mass
$M_{200}=2.7\times 10^9\, M_\odot$ and $c=r_{200}/r_s=14$. The
isothermal profile has $\rho_0=3\times 10^7 M_\odot/$kpc$^3$ and
$r_{\rm core}=1$ kpc.

The contours in the middle panel of Fig.~\ref{FigHalos} indicate the
constraints derived by \citet{Fattahi2016} on $M(<1$ kpc$)$. For
comparison, we also indicate with crosses the constraints at various
radii from \citet{Walker2011} and \citet{Amorisco2013}. The grey
shaded band corresponds to the results of the recent kinematic
analysis of Fornax's stellar component of \citet{Read2019}. Note how
all of these estimates concur at $r\sim 1$ kpc to a mass close to what
is assumed in our models.

For reference, the circular orbit timescale is
$t_{\rm circ}\approx 3\times10^8$ yr at $r=1$ kpc for both models; at
$r=0.1$ kpc, $t_{\rm circ}=8\times 10^7$ yr for the NFW case, and
$t_{\rm circ}=2.2\times 10^8$ yr for the cored halo.

\subsection{GC models}

GCs are modeled as single softened point masses. Three different
masses were chosen in our runs: a fiducial value of
$M_{\rm GC}=3\times 10^5\, M_\odot$, similar to Fornax GC3 (NGC 1049),
the most massive cluster orbiting Fornax \citep{Mackey2003a}. We also
explored models with $M_{\rm GC}=10^5\, M_\odot$, comparable to GC2,
GC4 and GC5. The other GCs in Fornax has much lower mass (GC1,
$3.7\times 10^4\, M_\odot$). Recall that dynamical friction times
scale inversely with mass. In the absence of other complicating
factors, and in the regime where the GC mass is small compared to that
enclosed within its orbit, the orbital decay of different clusters
should be similar, once their times are inversely scaled by cluster
mass. We assume that GC masses remain constant during the
evolution. This neglects possible mass losses due to internal
collisional processes within the cluster. Including this effect would
result in even longer orbital decay timescales than the ones reported
here, so our results may be regarded as conservative from that point
of view.

\subsection{N-body models}

Equilibrium N-body models with $1.6$ and $16$ million particles are
generated for each halo using the software package {\tt
  Zeno}\footnote{\tt http://www.ifa.hawaii.edu/faculty/barnes/zeno/}
developed by Josh Barnes at the University of Hawaii. This package
allows for the creation of a number of systems in virial equilibrium
by MonteCarlo sampling the appropriate distribution function.

The simulations were run with the publicly available {\tt Gadget2}
code \citep{Springel2005}, with standard numerical integration parameters. Pairwise
interactions between N-body particles are softened with a
Plummer-equivalent softening length of $\epsilon_{\rm P}=66.4$ and $210$ pc,
for the $16$M and $1.6$M particle halos, respectively. The
halo particle mass  is $1.78\times 10^2\, M_\odot$ (cusp) and
$1.99\times 10^2\, M_\odot$ (core) for
the 16M particle realizations. Particle masses are $10\times$ larger
for the 1.6M-particle halos. 

Each halo model is run for $\sim 4$ Gyr in isolation to allow them to
equilibrate and fully relax before introducing the GC. The profiles shown in
Fig.~\ref{FigHalos} are measured at the end of these
equilibration runs. Careful centering is required to obtain robust results; we use
in our analysis the reference frame given by the
gravitational potential-weighted center of all halo particles;
i.e., $\vec x_{\rm C}=\sum \Phi_i \vec x_i/\sum \Phi_i$; $\vec {\rm
  v}_{\rm C}=\sum \Phi_i \vec {\rm v}_i/\sum \Phi_i$

GC particles are softened with $\epsilon_{\rm P,GC}=13$ pc and are
introduced at the end of the equilibration period. They are placed at
various radii (typically $r_{\rm init}=0.5$, $1$, and $2$ kpc) on
circular orbits with random orientations. Their radial evolution is
then monitored as a function of time. Most of the runs reported here
correspond to the 1.6M model; a representative sample of those have
been repeated with the 16M-particle model, with indistinguishable
results. We have also repeated several runs varying
$\epsilon_{\rm P,GC}$. No significant variations were seen in the GC
orbital evolution for values of $\epsilon_{\rm P,GC}$ smaller than
adopted for our runs, although substantially longer dynamical friction
decay times were seen for (unrealistically) large values of
$\epsilon_{\rm P,GC}$. For $\epsilon_{\rm P,GC}\sim 10$ pc, for
example, GCs take roughly twice as long to decay than for our fiducial
value of $13$ pc.

\begin{figure}
\includegraphics[width=\columnwidth]{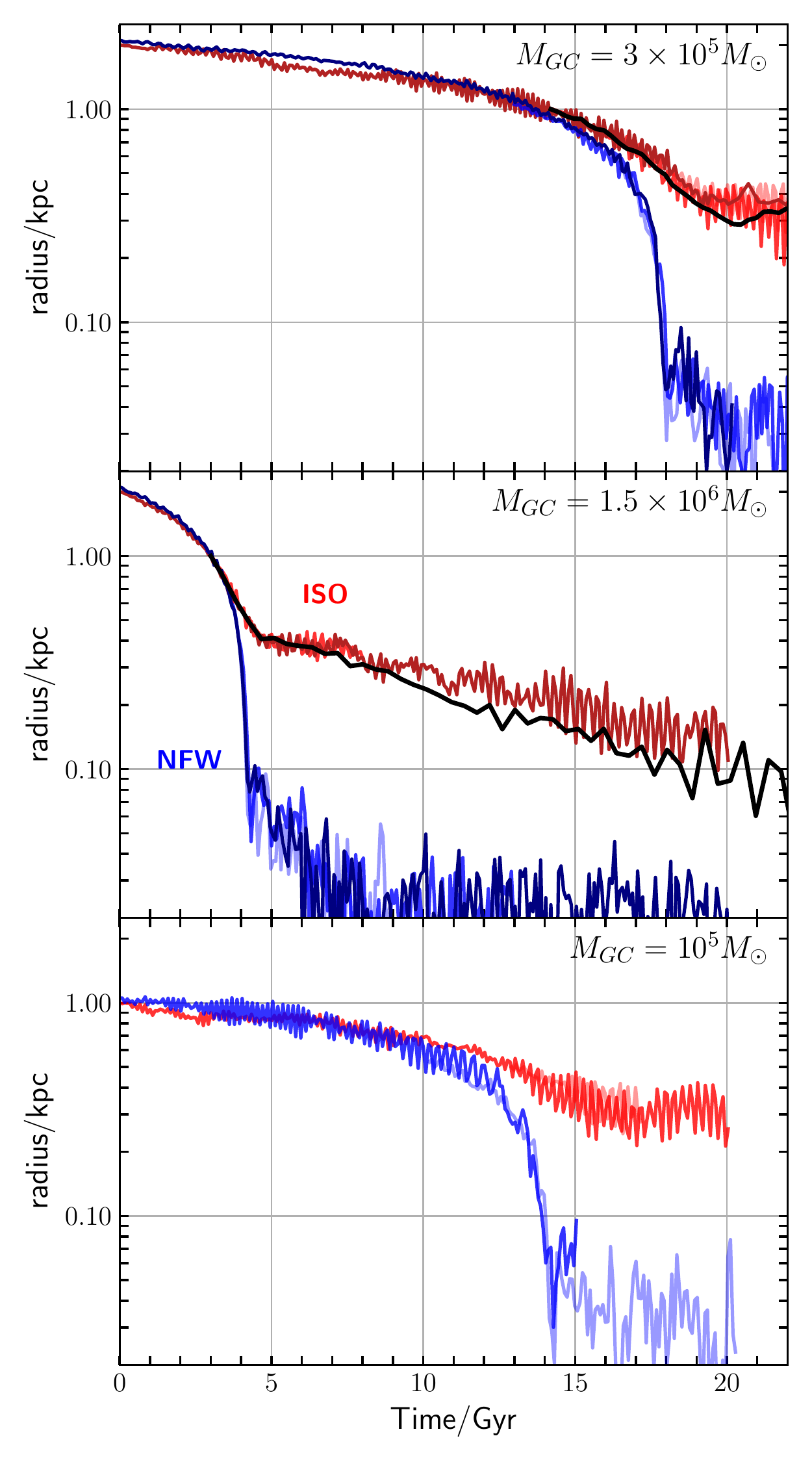}
\caption{Evolution of the radial distance of a
  $M_{\rm GC}=3\times 10^5\, M_\odot$ (top),
  $M_{\rm GC}=1.5\times 10^6\, M_\odot$ (middle), and
  $M_{\rm GC}= 10^5\, M_\odot$ (bottom) globular cluster. The
  evolution is followed for roughly $20$ Gyrs. The cuspy, NFW halo
  case is shown in blue; the core case in red. (The curves in black
  correspond to the 16M-particle halo model.) Different hues
  correspond to independent runs with different initial radii,
  $r_{\rm init}=2$, $1$, and $0.5$ kpc, respectively, and are shown
  after shifting their time origin so that their starting radii
  coincide. The near perfect overlap between different curves shows
  that the numerical results are independent of starting radii, as
  expected if clusters remain on a nearly circular orbits as they
  decay. Clusters either sink to the center (cusp) or stall (core),
  but do so on similar timescales. The top panel corresponds to a
  cluster with mass comparable to the most massive GC in Fornax
  (GC3/NGC 1049). Its orbit decays from $2$ to $1$ kpc in $\sim 13$
  Gyr, before stalling (core) or sinking (cusp) after $\sim 18$
  Gyr. The middle panel represents a cluster $5\times$ more massive
  than NGC 1049.  The bottom panel corresponds to a mass comparable to
  GC2, GC4 and GC5. }
    \label{FigRvsT}
\end{figure}

\section{Results}
\label{SecRes}

\subsection{Orbital decay timescales}

The time evolution of the fiducial mass GC
($M_{\rm GC}=3\times 10^5\, M_\odot$, similar to the most massive
Fornax GC, NGC 1049) is shown in the top panel of
Fig.~\ref{FigRvsT}. The figure shows the evolution of three different
runs per halo, each with different starting radii, $r_{\rm init}=2$,
$1$, and $0.5$ kpc. Curves for the latter two have been shifted
horizontally so that they coincide in radius and time, at the beginning, with the
$r_{\rm init}=2$ kpc case. All three curves are essentially
indistinguishable from each other. This highlights the fact that the
GC evolution is independent of starting radius, as expected if
orbits remain roughly circular throughout the evolution.

This figure illustrates a few interesting points. One is that, if NGC
1049 had formed at $2$ kpc from the center, then it would only have
decayed to a distance of $\sim 1$ kpc after a Hubble time.  The
orbital decay accelerates once the cluster reaches $1$ kpc, and the
cluster quickly sinks to the center in the case of the cusp, or
``stalls'' at $r_{\rm stall}\sim 0.3 \, r_{\rm core}=300$ pc in the
case of the core\footnote{Note that GCs do not truly stall at
  $r_{\rm stall}$; rather, the rate of their inspiraling slows down
  when clusters reach that radius. See Sec.~\ref{SecLongTerm} below
  for details.}.

This behaviour is consistent with earlier work \citep[see;
e.g.,][]{Goerdt2006,Read2006,Cole2012}: GCs always stall at
$\sim 0.3\, r_{\rm core}$, when the core radius is defined as that
where the density drops to half its central value.

Interestingly, the time the cluster takes to either sink or stall is
approximately the same, $\sim 18$ Gyrs ($\sim 4$ Gyrs since the
cluster reached $1$ kpc) in both cases. In other words, {\it dynamical
  friction timescales in cored or cuspy halos are essentially
  indistinguishable} for halos normalized as in
Fig.~\ref{FigHalos}. The difference is in the final radius reached by
the cluster: $\sim 300$ pc in the case of the core, or the center in
the case of the cusp.

The middle panel of Fig.~\ref{FigRvsT} confirms this conclusion for the case of a
cluster $5\times$ more massive, $M_{\rm GC}=1.5\times 10^6\,
M_\odot$. The evolution of this cluster is exactly analogous to that
of its less massive counterpart shown in the  top panel. The
timescales to sink or stall are still roughly the same for cusp or
core, albeit $5\times$ shorter than in the former case, just as expected
from the mass ratio between those clusters.

Conversely, for clusters less massive than our fiducial mass, the
decay timescales are substantially longer. The results for
$M_{\rm GC}=10^5\, M_\odot$ (comparable to Fornax GC2, GC4, and GC5)
are shown in the bottom panel of Fig.~\ref{FigRvsT}, and show that
clusters with $r_{\rm init}=1$ kpc take more than $13$ Gyrs to either
sink or stall. This is as expected from the fiducial
($3\times10^5\, M_\odot$) mass case, which takes $\sim 4$ Gyrs to sink
or stall from a radius of $1$ kpc.  Placed at
$r_{\rm init}>1$ kpc, GC2, GC4, and GC5 would have barely evolved
over $13$ Gyr. Again, the evolution shown in the three panels of
Figs.~\ref{FigRvsT} are all analogous and consistent with each other,
once times are scaled by the mass of a cluster and comparisons are
made for the same starting radius.

\begin{figure}
\includegraphics[width=\columnwidth]{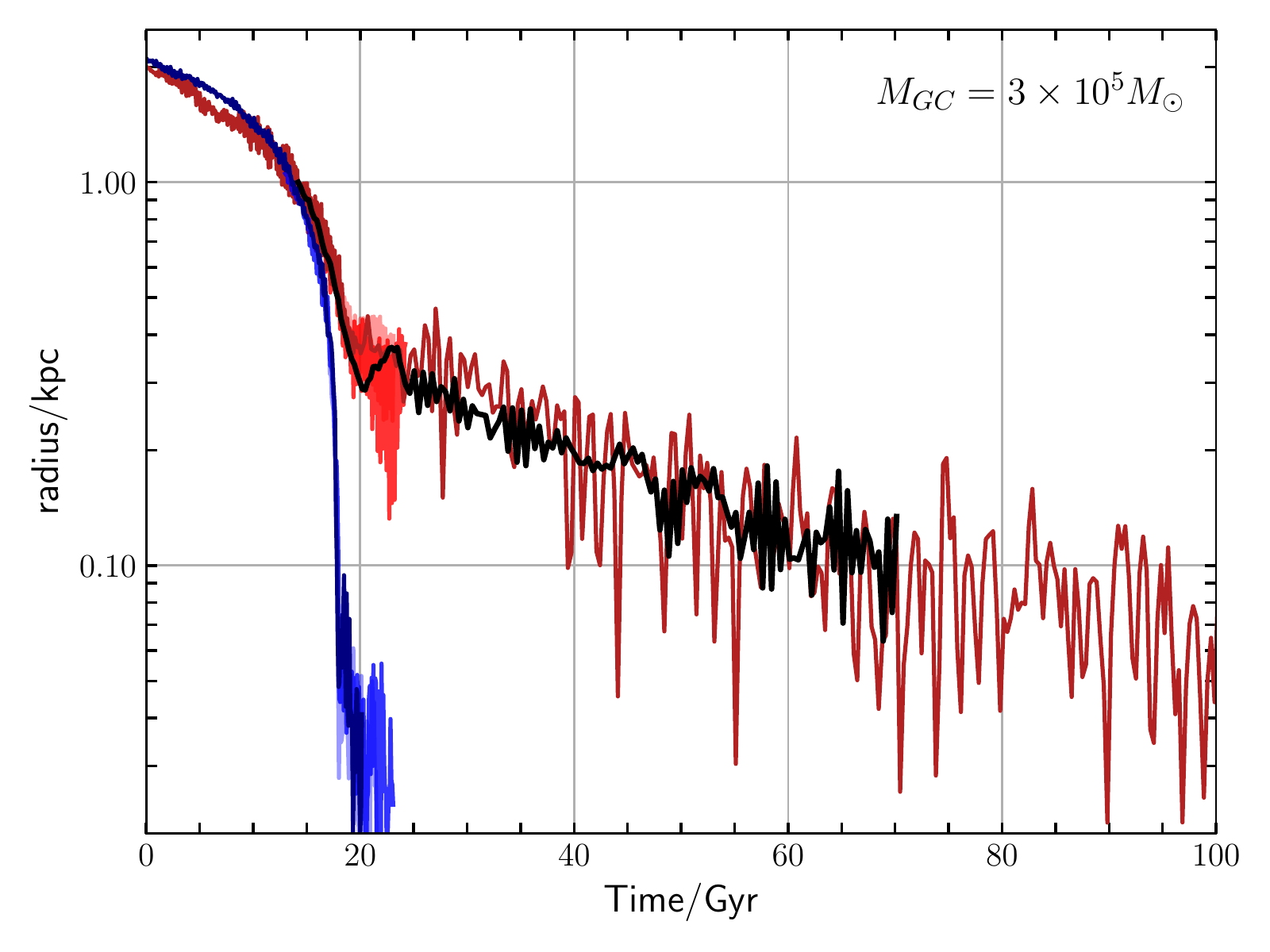}
\caption{As Fig.~\ref{FigRvsT}, but following the evolution for
  $t=100$ Gyr. Note that the GC keeps decaying inside the core, but on
  a $\sim 5\times$ longer timescale than in Fig.~\ref{FigRvsT}, as
  expected given the mass ratio between the clusters. }
    \label{FigRvsT100Gyr}
\end{figure}

\begin{figure}
\includegraphics[width=\columnwidth]{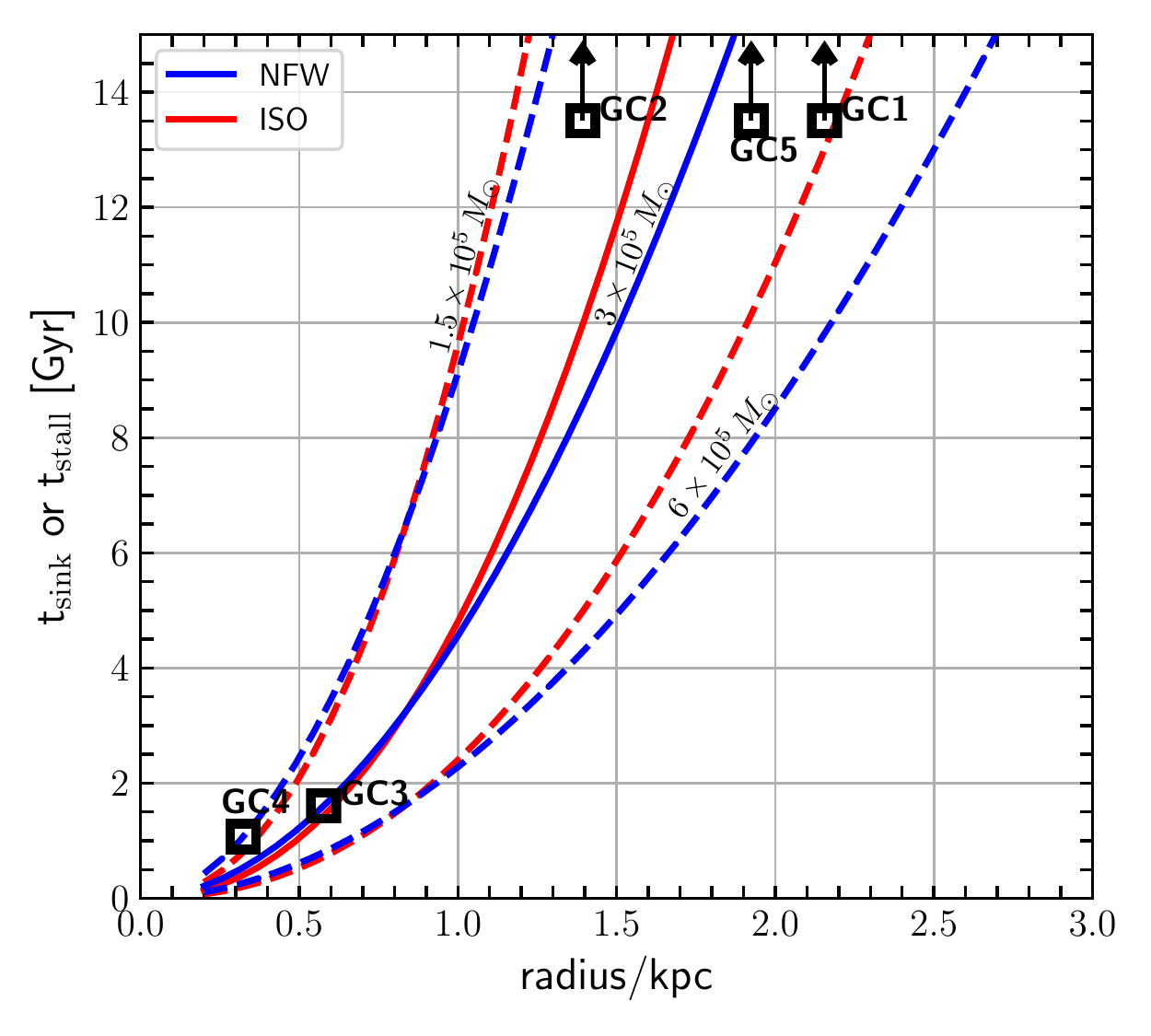}
\caption{Time to stall (core, in red) or sink (cusp, in blue) as a
  function of GC mass and initial radius. Note that, at given radius
  and mass, the timescales are similar for cuspy and cored halos
  normalized to the same mass within $r_{\rm core}=1$ kpc
  (Fig.~\ref{FigHalos}). As expected, timescales scale inversely with
  mass, and are strongly dependent on initial radius. Fornax GCs are
  placed on this figure at radii equal to $4/3 \times$ the
  present-day projected distance, and at a location consistent with
  its mass. Note that only GC3 and GC4 are expected to evolve
  significantly due to dynamical friction over the next few Gyrs. See
  text for a full discussion}
    \label{FigTsRinit}
\end{figure}

\subsection{Long-term evolution}
\label{SecLongTerm}

The middle panel of Fig.~\ref{FigRvsT} follows the evolution of the
most massive cluster in our series for $\sim 20$ Gyr, quite a long
period of time after its initial stall/sink. This allows us to probe
the long-term evolution of the clusters once they reach the inner
regions of the halo. In the case of the cusp, once the cluster sinks
to the center it stays there. In the case of the core, after its
initial stall the cluster keeps losing energy and slowly drops deeper
inside the core.  At the end of the simulation the cluster has reached
a radius of $\sim 200$ pc, roughly where the halo enclosed mass is
comparable to its own (see Fig.~\ref{FigHalos}). Note that we find the
same result for the 1.6M and 16M-particle halos, so the long-term
sinking behaviour seems robust.

This long-term evolution is not unique to this massive cluster. The
fiducial mass GC also keeps losing energy after its initial stall, as
shown in Fig.~\ref{FigRvsT100Gyr}. The main difference is that this
long-term trend takes, as expected, $5\times$ longer, and is therefore
only noticeable in simulations that follow the evolution for roughly
$\sim 100$ Gyr. Indeed, after that time the cluster has shrunk its
orbit to roughly $100$ pc, which is about the radius where the
enclosed mass of the halo matches that of the cluster.

These results seem to disagree with those of \citet{Cole2012}, who
report that clusters that drop deep into the core of a halo are pushed
out by a mechanism they call ``dynamical buoyancy''.  This effect was
only seen in the case of their ``large core'' (LC) halo, which is
actually quite similar to the ISO halo we adopt here. Indeed, the LC
halo has $r_{\rm core}\sim 1.2$ kpc (only $20\%$ larger than ISO's)
and $\rho_0\sim 4\times 10^7\, M_\odot$/kpc$^3$ (about $33\%$ larger
than ISO's). The main difference is that the LC density profile
steepens faster than ISO's: at $r_{\rm core}$ the logarithmic slope is
$d\log\rho/d\log r=-1.8$ for LC and $-1.1$ for ISO. This difference
seems, at face value, too small to explain why we do not see
``dynamical buoyancy'' in our runs. At this point it is unclear what
the origin of the discrepancy might be, but it is something that we
plan to investigate in future work.

\section{Discussion and Conclusions}
\label{SecConc}

In agreement with earlier work, our simulations indicate that the GC
population of Fornax is expected to evolve continuously due to
dynamical friction. For the halo models considered here
(Sec.~\ref{SecHaloModels}), our results are summarized in
Fig.~\ref{FigTsRinit}, where we plot the time it would take for
clusters of various masses to sink (NFW, blue) or stall (ISO,
red). These times are computed by fitting the results of our
simulations with simple power laws and, therefore, in a strict sense,
are rough estimates that apply only to clusters in circular orbits. However, these times are
not expected to differ much from those for clusters in non-circular
orbits with comparable average radii \citep{Angus2009}.

We compare these results with Fornax GCs (shown with open squares in
Fig.~\ref{FigTsRinit}), taking into account the masses of individual
clusters and assuming that they are at radii $4/3\times$ their current
projected distance. These results indicate that GC1, GC2, and GC5 are
either too far, or have too little mass, to decay significantly, even
over a timespan as long as the next $10$-$15$ Gyr. It is thus highly
unlikely that all clusters are today at a common radius dictated by
dynamical friction effects.

On the other hand, both GC3 and GC4 \citep[the two closest to the
center, with projected distances of $0.43$ and $0.24$ kpc,
respectively;][]{Mackey2003a} should either sink or stall over the
next few Gyr, according to Fig.~\ref{FigTsRinit}.  Could it be that
Fornax has a core and these two clusters have ``stalled'' at a common
radius?  This possibility may in principle be checked using the radial
velocities of these clusters relative to Fornax. GC4, in particular,
has a well defined radial velocity offset of nearly $\sim 10$ km/s
relative to Fornax \citep{Hendricks2016}. This is significantly higher
than the expected circular velocity at its present deprojected radius
(in the case of a core), so it is highly unlikely that this cluster is
actually close to its stalling radius.

This leaves GC3, which, if ``stalled'', would imply
$r_{\rm stall}\sim 600$ pc, its inferred 3D distance from the
center\footnote{ There is, of course, also the possibility that this
  cluster is much further away in distance and lie, by chance, only in
  projection near the center of Fornax. This would make the case for a
  core even weaker, and could be checked by inspecting the relative
  proper motion of GC3 relative to Fornax, an issue we are currently
  working on.}. This implies $r_{\rm core} \sim 2$ kpc (i.e., at
least twice as large as its stellar half-mass radius; recall that
$r_{\rm stall}\approx 0.3\, r_{\rm core}$).

A core radius this large seems difficult to accommodate in either of the
two leading scenarios for core creation; i.e., baryonic
outflow-induced cores, or dark matter self-interactions. Indeed, if
cores are carved out of CDM halos through stellar feedback, then it
would be difficult to explain a core size at least twice as large as
the half-light radius of the galaxy \citep{Pontzen2014,Oman2016}.

On the other hand, if cores are due to self-interacting dark matter,
these would be expected to be of sub-kpc scale in galaxies as small as
Fornax, even for extreme values of the cross
section. \citet{Elbert2015}, for example, report sub-kpc core
radii\footnote{Recall that our definition of core radius follows the
  traditional convention of designating the distance where the density
  drops by a factor of two from the central value.} even for halos
substantially more massive than Fornax, and for all values of the
cross section in the plausible range of $0.1$-$1$ cm$^2$/g \citep[the
same is true even for larger cross sections; see,
e.g.,][]{Sameie2019}.

If, on the other hand, Fornax has a cusp, then GC3 and GC4 must be on
their way to sinking to the center, having started their decay
from $r_{\rm init}\sim 1.5$ kpc (GC3) and $r_{\rm init}\sim 1$ (GC4)
about $\sim 10$ Gyr ago \citep[their typical ages; see,
e.g.,][]{Buonanno1998}. These initial radii are quite plausible, as
they lie well within the inferred tidal radius of Fornax imposed by
the Galactic tides, which is estimated to be of order $1.8$-$2.8$ kpc
\citep{Angus2009,Cole2012}.

It could be argued that, because the sinking accelerates once GCs
reach the inner regions of the halo, this represents a ``fine-tuning''
problem. In other words, why are we observing GC3 and GC4 at such
radii and not at the center if they are at a rapidly evolving stage of
their decay?  The same fine-tuning argument may be used against a
core, however, since in that case GCs also accelerate their decay
before stalling, and the timescales to sink or stall are very
similar. This argument only favours a core if both clusters have
stalled, which, as discussed above, is disfavoured by the radial
velocity of GC4 and requires an implausibly large core radius of at
least $2$ kpc.

In the case of the cusp, the disadvantage of a scenario where GC3 and
GC4 formed at slightly larger initial radii and are at present on
their way to sinking to the center is that all clusters would then
have formed outside the present-day half-light radius of the dwarf. In
the absence of a well-defined theory of GC formation it is difficult
to assess the severity of this objection, but it should be noted that
Fornax GCs are older and more metal-poor than most stars in the dwarf.
Such populations tend to be more spatially extended than younger and
more metal-rich ones, in Fornax \citep{Battaglia2006,Walker2011} as
well as in other dwarfs such as Sextans \citep{Battaglia2011} and
Sculptor \citep{Tolstoy2004}.

Some of these differences could indicate
an ancient merger, which would have dispersed the old stellar component and
allowed the enriched gas to sink further in before forming stars
\citep{Benitez-Llambay2016,Genina2019}. This would provide a plausible  
explanation for the radial offset between the original distribution of
Fornax GCs and the present-day distributions of its stars.

We end by noting that our simulations show no clear evidence of the
``dynamical buoyancy'' effects reported by \citet{Cole2012}. It is thus unclear at
this point what the origin of the difference might be, but it does
underscore the need for further study of the effect, including a
theoretical explanation and an exploration of its dependence on
cluster mass and on the detailed dynamical properties of the core.

The Fornax GC spatial distribution is thus unlikely to help discern between
cusp and core. In this sense, the GC timing problem is no different from
dynamical analyses that use the spherical Jeans' equations to derive
mass profiles from velocity dispersion and density profile data. These
models suffer from well-known degeneracies that prevent a conclusive
determination of the shape of the inner density profile \citep[see;
e.g., the reviews by][and references
therein]{Strigari2013,Walker2013}. Indeed, data for several dSphs are
consistent with NFW cusps {\it and} cores
\citep[e.g.,][]{Gilmore2007,Strigari2010}.

Using higher-order moments of the line-of-sight velocity distribution
offers in some cases the possibility of breaking the
degeneracy. Recently, \citet{Read2019} applied this method to Fornax
and concluded that the dark matter density drops by about an order of
magnitude (from $\sim 10^8$ to $10^7\, M_\odot/$kpc$^3$) over the
range $0.1$ to $1$ kpc (see the middle panel of their Fig.~3). This is
close to what is expected for a $\rho\propto r^{-1}$ NFW cusp and is
only slightly less concentrated than the model we analyze here (see
the middle panel of Fig.~\ref{FigHalos}).

Our overall conclusion is that it is unclear how or whether the
spatial distribution of GCs in Fornax may be used to discern between
the core and cusp scenarios.  What is clear, though, is that it cannot
be used to argue convincingly against the presence of a cusp in the
inner density profile of the Fornax dwarf spheroidal.

\section*{Acknowledgements}

JFN acknowledges useful discussions with Justin Read and Nelson
Caldwell. We also thank Christian Johnson for bringing to our
attention the work of \citet{Hendricks2016}.



\bibliographystyle{mnras}
\bibliography{refs} 

\end{document}